\documentclass[aip,reprint]{revtex4-1}

\usepackage{mathtools}

\usepackage{enumitem}
\usepackage{xr-hyper}
\usepackage{hyperref}
        \hypersetup{
           breaklinks=true,   
           colorlinks=false,   
           pdfusetitle=true,  
        }

\usepackage{lineno}
\usepackage{siunitx}
\usepackage{amsmath}
\usepackage{amssymb}
\usepackage{graphicx}
\usepackage[T1]{fontenc}
\usepackage{todonotes}
\usepackage{gensymb}

\usepackage[utf8]{inputenc}

\begin{document}


\title{Broadband surface-emitting THz laser frequency combs with inverse-designed integrated reflectors}
\vspace{2 cm}

\author{Urban Senica}
 \email{usenica@phys.ethz.ch}
\author{Sebastian Gloor}
\author{Paolo Micheletti}
\author{David Stark}

\author{Mattias Beck}
\author{J{\'e}r{\^o}me  Faist}
\author{Giacomo Scalari}
 \email{scalari@phys.ethz.ch}

\affiliation{
Quantum Optoelectronics Group, Institute of Quantum Electronics, ETH Z{\"u}rich, 8093 Z{\"u}rich, Switzerland
}


\date{\today}

\begin{abstract}
THz quantum cascade lasers (QCLs) based on double metal waveguides feature broadband and high-temperature devices for use in spectroscopy and sensing. However, their extreme field confinement produces poor output coupling efficiencies and divergent far-fields.
Here, we present a planarized THz QCL with an inverse-designed end facet reflector coupled to a surface-emitting patch array antenna. All the components have been optimized for octave-spanning spectral bandwidths between 2-4 THz and monolithically integrated on the same photonic chip. We demonstrate this experimentally on broadband THz QCL frequency combs, with measured devices showing a seven-fold improvement in slope efficiency compared to devices with a cleaved facet. They feature peak powers of up to 13.5  mW with surface emission into a narrow beam with a divergence of (17.0\degree x 18.5\degree), while broadband fundamental and harmonic comb states spanning up to 800 GHz are  observed.

\end{abstract}




\maketitle
 
\section*{Introduction}
THz quantum cascade laser (QCL)\cite{kohler_terahertz_2002} frequency combs \cite{burghoff_terahertz_2014} and dual combs \cite{rosch_-chip_2016} are compact sources of coherent THz radiation, promising for use in broadband spectroscopy and sensing.

However, a major practical limitation has been their low output powers and poor far-field patterns. Both properties originate from the double metal waveguide cavity configuration. Since the propagating optical mode is confined to extremely subwavelength dimensions (typically, $d=\sim\SI{10}{\micro\metre}$ for a central emission wavelength of $\lambda_0 =\sim\SI{100}{\micro\metre}$), it acts as a point-like source and produces highly divergent and frequency-dependent far-field patterns. Additionally, due to metallic waveguide confinement and the resulting large impedance mismatch between the guided and free space optical mode, the facet reflectivities are relatively high, in the order of $R=70\%$ at a frequency of 3 THz. While this does reduce the mirror losses, it also limits the slope efficiencies and output powers.
There have been a variety of approaches to improve the outcoupling properties of THz QCLs, but these have either been optimized for narrowband emission\cite{amanti2009low, Bosco2016}, have an intrinsically limited bandwidth\cite{rosch2017broadband, Xu2012}, or require additional post-processing and mounting steps\cite{lee2007high, Senica2020}. 

\begin{figure*}[htbp]
\centering
\includegraphics[width=0.8\linewidth]{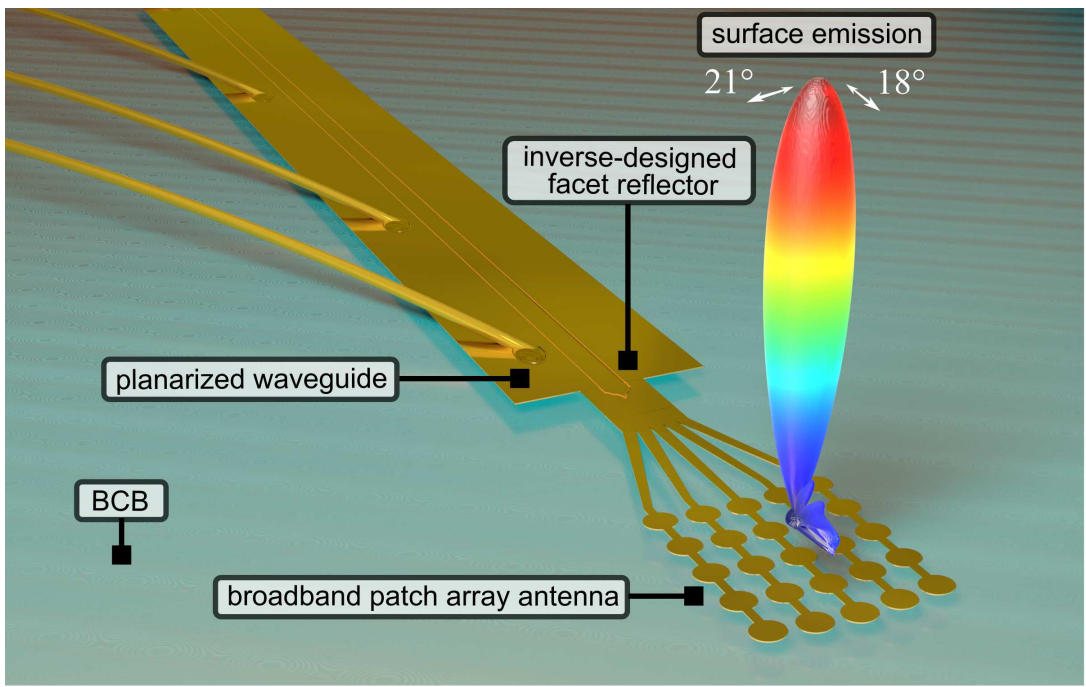}

\caption{Illustration of the fabricated device: on the front side of a planarized ridge waveguide is an inverse-designed end facet reflector, which is coupled to a passive waveguide connecting to a broadband patch array antenna for surface emission. A spherical plot of the simulated broadband far-field pattern is superimposed on top of the antenna structure (same as in Fig. \ref{fig:farfield}(b), where the measured emission spectrum is included in the far-field simulation).  The emitted light is linearly polarized (aligned with the ridge waveguide).}
\label{fig:structure}
\end{figure*}

Here, we use inverse-designed end facet reflectors for precise control of the mirror losses and output power, and couple the optical mode to a surface-emitting patch array antenna, all monolithically integrated on the same photonic chip, as illustrated in Fig. \ref{fig:structure}. All the designed components have been optimized for an octave-spanning emission spectrum between 2-4 THz and improve the output power and far-field pattern while featuring broadband comb states simultaneously.

\section*{Results}

\subsection*{Inverse-designed end facet reflectors}
We first address the limited output powers due to the high reflectivity of cleaved end facets of double metal waveguides\cite{Kohen_doublemetal_2007}. In our recent work, we have shown that the planarized waveguide platform enables the reduction of end facet reflectivities by coupling the active waveguide into a passive waveguide, defined with a metallic stripe on top of the surrounding low-loss polymer material with a lower refractive index (BCB, n = 1.57) \cite{senica2022planarized}. In the simplest configuration, a flat dry-etched planarized facet results in a reduced reflectivity of around R = 23\% at a frequency of 3 THz into the passive waveguide. This is already a remarkable reduction with respect to the cleaved facet value of $70\%$. 

In order to have precise control of the facet reflectivity over a broad bandwidth, we implemented an inverse design approach based on adjoint optimization \cite{Lalau-Keraly2013}. In recent years, inverse design has emerged as a powerful design and optimization tool in various areas and applications in photonics \cite{molesky2018inverse}. 
The main advantage of such an approach is that instead of manual parameter sweeps and fine-tuning to achieve a high-performance device design, an optimization algorithm is used to automatically adapt the structure in an iterative loop to maximize the desired figure of merit without any user input. 

In our specific geometry, we implemented an inverse-design shape optimization simulation loop, where the outline shape of the end facet is modified to match the desired reflectivity value. The details of the parametrization and implementation can be found in the Supplemental Material. Due to the symmetric planarized waveguide structure, only a 2D slice of the structure and an in-plane (x-y, perpendicular to the growth direction z of the heterostructure) propagation simulation are required. Additionally, the adjoint optimization approach features very fast convergence, as it only requires two simulations of the current structure geometry (a forward and a backward propagation simulation) to compute the gradients and update all the geometrical parameters in a single step. These favourable aspects result in a very efficient optimization routine which produces optimized designs after only around 25 iterations, taking less than an hour on a normal desktop computer.

\begin{figure}[htbp]
\centering
\includegraphics[width=1\linewidth]{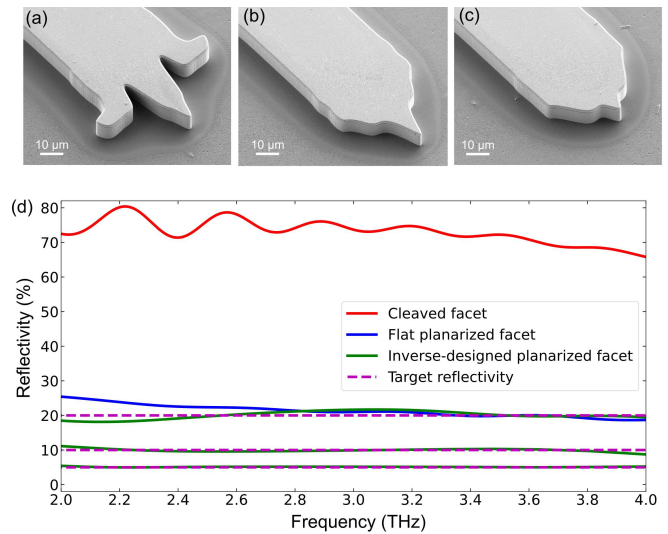}
\caption{\textbf{(a-c)} SEM images of the dry-etched inverse design facet reflectors with designed facet reflectivities of 5\%, 10\% and 20\%, respectively. Subsequently, the waveguides are planarized with BCB and an extended top metallization and antenna extractor are deposited. \textbf{(d)} Simulated reflectivities of a cleaved facet, a flat planarized facet, and the inverse-designed planarized facets, where the simulated reflectivities match well with the target reflectivities.}
\label{fig:facets}
\end{figure}

In Fig. \ref{fig:facets}(a-c), we show SEM images of fabricated dry-etched inverse-designed end facets, designed to have a 5\%, 10\% 
 and 20\% reflectivity over octave-spanning spectra between 2-4 THz. Subsequently, the active waveguides are planarized with a low-loss polymer BCB and the top metallization is defined (see Ref. \cite{senica2022planarized} for details). 
 A comparison between simulated reflectivities of cleaved, flat planarized, and inverse-designed planarized facets is shown in Fig. \ref{fig:facets}(d). The facets do not induce any considerable group delay dispersion after reflection, which could otherwise be detrimental to comb formation.

\subsection*{Surface-emitting patch array antenna}
After a partial reflection at the planarized end facet, the optical mode is guided within a passive waveguide and coupled into a surface-emitting patch array antenna. As the propagating light wave spreads out into the antenna branches, the individual patches are oscillating in phase and combine into a narrow vertical beam. The basic design intended for single-mode operation was presented in Ref. \cite{Bosco2016}. 

\begin{figure}[htbp]
\centering
\includegraphics[width=1\linewidth]{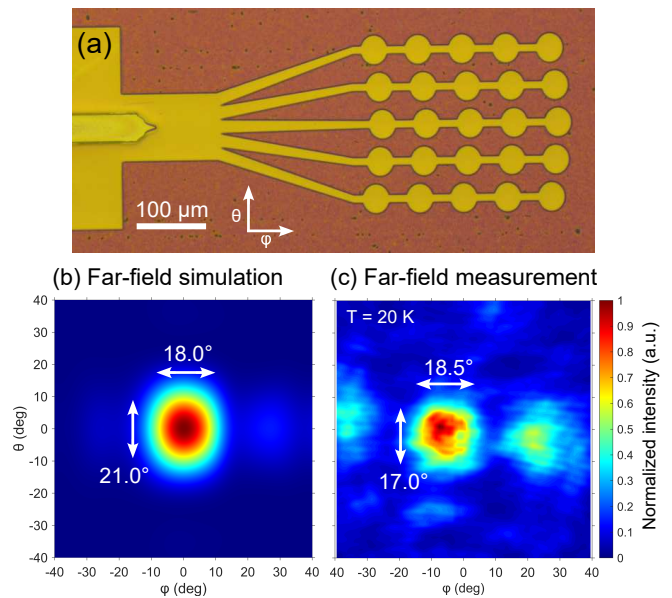}
\caption{\textbf{(a)} Optical microscope image of the front side of a planarized ridge waveguide, where an inverse-designed facet reflector is coupled via a passive waveguide into a broadband patch array antenna for surface emission. \textbf{(b)} Broadband far-field simulation, which uses the emission spectrum of the measured device, produces a single-lobed pattern. \textbf{(c)} Far-field measurement of a broadband device agrees well with the simulation with a FWHM beam width of (17.0\degree x 18.5\degree).}
\label{fig:farfield}
\end{figure}

Here, we developed a broadband patch array antenna, shown in the optical microscope image in Fig. \ref{fig:farfield}(a). This design has also been optimized for octave-spanning emission spectra, matching the same broad frequency range as the reflector structures. 
First, to reduce the beam divergence, the antenna emission area was enlarged by increasing the number of patch elements to (5x5). For broadband emission, the optimal shape, size and positioning of the individual antenna elements were found with full-wave numerical simulations of the surface emission, where a minimal beam divergence and beam steering with frequency were obtained.


In Fig. \ref{fig:farfield}(b) we show the simulated broadband far-field pattern which features a single narrow lobe in the vertical emission direction. The result was obtained by including the emission spectrum of the measured device with frequency-dependent far-field simulation results. Specifically, this was done with a spectrally-weighted linear sum of simulated far-field patterns, as described in more detail in Ref.\cite{Senica2020}.

The antenna far-field measurement in Fig. \ref{fig:farfield}(c) agrees well with the simulations with a full-width half-maximum (FWHM) beam divergence of (17.0\degree x 18.5\degree). The measurement was performed on a broadband emission sample using a pyroelectric detector (Gentec-EO: THZ2I-BL-BNC) mounted on a motorized angular scanning stage. The laser was driven in micro-pulse (500-ns-long pulses, 20\% duty cycle), macro-pulse mode (30 Hz, 50\% duty cycle), where the emission spectrum was spanning between around 2.3-3.3 THz.

\subsection*{High-power surface emission}
With a reduction of the front end facet reflectivity $R_1$, the mirror losses are increased through the relation:
\begin{equation}
\label{eq:loss_mirror}
\begin{aligned}
    \alpha_\mathrm{m}=-\frac{1}{2L} \ln(R_1R_2)
\end{aligned}
\end{equation}

where $L$ is the waveguide length and $R_2$ the back mirror reflectivity. On the one hand, together with waveguide losses $\alpha_{\mathrm{wg}}$, this increases the total losses $\alpha_{\mathrm{tot}}=\alpha_{\mathrm{wg}}+\alpha_{\mathrm{m}}$ with an expected increase of the lasing threshold. On the other hand, the slope efficiency\cite{faist_quantum_2013} should be increased:
\begin{align}
\label{eq:slope_efficiency}
    \frac{\mathrm{d}P}{\mathrm{d}I}=\frac{N_{\mathrm{p}} h \nu}{e}\frac{\alpha_{\mathrm{m,1}}}{\alpha_{\mathrm{tot}}}\frac{\tau_{\mathrm{eff}}}{\tau_{\mathrm{eff}}+\tau_{\mathrm{2}}}
\end{align}
When comparing different devices fabricated using the same active region, only the $\frac{\alpha_{\mathrm{m,1}}}{\alpha_{\mathrm{tot}}}$ ratio is changing. 

To evaluate the performance of the inverse-designed end facet reflectors, we experimentally compare three different front (extracting) facets of devices processed on the same chip. These are a cleaved facet, a flat planarized facet with the antenna, and a 10\% reflectivity inverse-designed planarized facet with the antenna.
The first one has a high-reflectivity back facet and a total length of 2.8 mm, while both antenna samples have a cleaved back facet and a length of 2.5 mm. All the samples have the same active waveguide width of \SI{40}{\micro\metre} to ensure the same waveguide loss $\alpha_{\mathrm{wg}}$ and nearly the same total active device area (difference of <5\%).

\begin{figure}[htbp]
\centering
\includegraphics[width=1\linewidth]{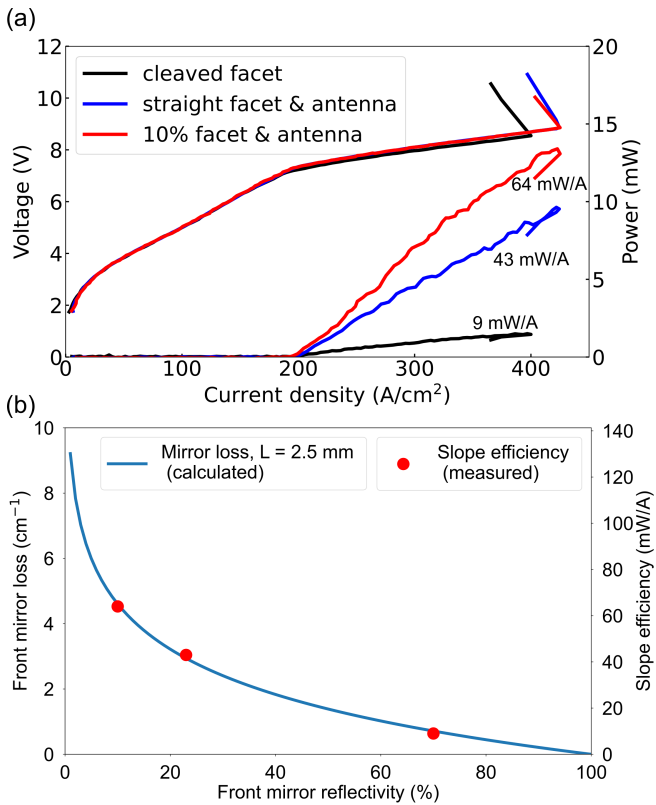}
\caption{\textbf{(a)} LJV curves comparison of devices with a cleaved facet, a flat planarized facet and a 10\% reflectivity inverse-designed facet. While the threshold current density increases only slightly, the output power and slope efficiency increase significantly with the planarized facets and an antenna. \textbf{(b)} Calculated front mirror loss for the different measured facets (blue line is for a cavity length of 2.5 mm). The red dots mark the measured slope efficiencies, which follow the same dependence with a varying front mirror reflectivity.}
\label{fig:liv}
\end{figure}

In Fig. \ref{fig:liv}(a), we plot the measured LJV curves of the three devices, all characterized under the same operating conditions (heat sink temperature of 20 K, 500 ns pulses and a duty cycle of 20\%). The output power was measured by a large area  calibrated absolute power meter (Thomas Keating Ltd.), which ensures the whole THz emission is collected (even in the case of poor far-field patterns of cleaved facets). When measuring the THz emission within a narrower range of spatial angles (typical for spectroscopy experiments), the improved collection efficiency due to the antenna would result in an even more favourable ratio of slope efficiencies. The comparison of calculated and experimental results for the three devices is summarized in Table \ref{tab:comparison}.

\renewcommand{\arraystretch}{1.2} 
\begin{table}[!ht]
    \centering
    \begin{tabular}{|c||c|c|c|}
    \hline
        Parameter & Cleaved facet & Flat facet & 10\% facet \\ \hline\hline
        $R_1$ (\%) & 70 & 23 & 10 \\ 
        Length (mm) & 2.8 & 2.5 & 2.5 \\
        $I_{\mathrm{thr}}$ (mA) & 178 & 195 & 195 \\
        $J_{\mathrm{thr}}$ (A/cm\textsuperscript{2}) & 185 & 195 & 195 \\
        $\alpha_{\mathrm{m, 1}}$ (cm\textsuperscript{-1}) & 0.64 & 2.94 & 4.61 \\ 
        $\alpha_{\mathrm{tot}}$ (cm\textsuperscript{-1}) & 20.6 & 23.7 & 25.3 \\
        $\frac{\mathrm{d}P}{\mathrm{d}I}$ (mW/A) & 9 & 43 & 64 \\  
        $P_{\mathrm{max}}$ (mW) & 1.5 & 9.5 & 13.5 \\  \hline\hline
        $\alpha_{\mathrm{m, 1}}$ ratio  & 1 & 4.6 & 7.2 \\ 
        $\frac{\mathrm{d}P}{\mathrm{d}I}$ ratio & 1 & 4.8 & 7.1 \\
        \hline
    \end{tabular}
    \caption{Summary of an experimental comparison of three different types of laser waveguide end facets.}
    \label{tab:comparison}
\end{table}

When replacing the front cleaved facet with a planarized antenna-coupled facet, the threshold current density $J_{\mathrm{thr}}$ only increases slightly (from 185 A/cm\textsuperscript{2} to 195 A/cm\textsuperscript{2}). This suggests that for such waveguide lengths ($L$ = 2.5 mm and longer), the total losses are dominated by waveguide losses.
Indeed, the waveguide losses of double metal waveguides are estimated to be in the order of 20 cm\textsuperscript{-1}(due to the overlap with lossy metals, intersubband absorption and scattering losses from sidewall roughness), while the computed front mirror losses are below 5 cm\textsuperscript{-1} for all the considered types of facets. 
In such a case, the comparison of slope efficiencies can be simplified from $\frac{\mathrm{d}P}{\mathrm{d}I}\propto\frac{\alpha_{\mathrm{m,1}}}{\alpha_{\mathrm{wg}}+\alpha_{\mathrm{m}}}$ to $\frac{\mathrm{d}P}{\mathrm{d}I}\propto\alpha_{\mathrm{m,1}}$. In Fig. \ref{fig:liv}(b), we plot the calculated front mirror loss (blue line) and the measured slope efficiencies (red dots). As these follow the same dependence versus the front mirror reflectivity, we are indeed in a regime where waveguide losses dominate over mirror losses. The measured slope efficiency of 64 mW/A of the sample with the 10\% inverse-designed facet and the antenna is a factor of 7.1 higher than for the cleaved facet reference sample, with a measured peak power of 13.5 mW. We should also note here that while reducing the reflectivity to even lower values (close to $R_1=0\%$) is tempting in terms of predicted slope efficiencies, the quickly increasing mirror losses would start to increase the laser threshold significantly, eventually preventing lasing.


\subsection*{Broadband frequency combs}
After demonstrating the improved far-field and outcoupling efficiency properties, we now highlight the broadband frequency comb performance. It is worth noting that a reduced cavity feedback has been predicted to be beneficial also for the comb operation itself\cite{humbard2022analytical, beiser2021engineering, senica2022planarized}.

\begin{figure}[htbp!]
\centering
\includegraphics[width=1\linewidth]{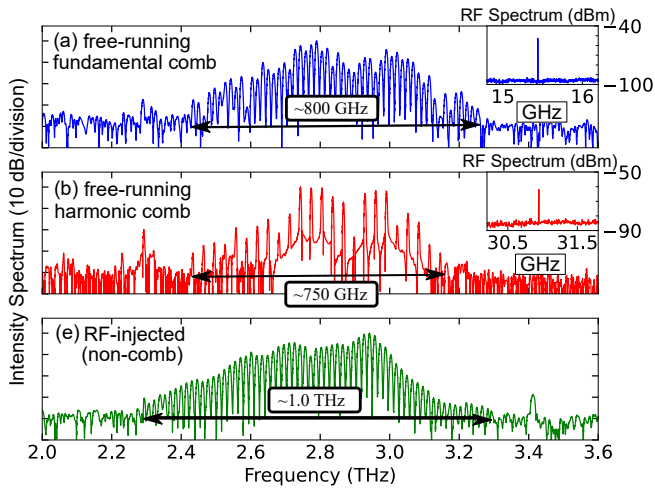}
\caption{Measured THz and RF spectrum of the 10\% reflectivity inverse-designed facet device in CW at a heat sink temperature of 30 K. \textbf{(a)} Free-running fundamental comb spanning 800 GHz and a strong single RF beatnote at the roundtrip frequency of $f_{\mathrm{rep}}$ = 15.4 GHz. \textbf{(b)} Free-running second harmonic comb spanning 750 GHz and a strong single RF beatnote at $2f_{\mathrm{rep}}$ = 30.8 GHz. \textbf{(c)} Strongly RF-injected (+32 dBm at source) incoherent state spanning around 1 THz.}
\label{fig:spectrum}
\end{figure}

In Fig. \ref{fig:spectrum}, we show the measured THz and RF spectra of the 10\% reflectivity inverse-designed facet device from the previous section (length $L$ = 2.5 mm, width $w$ = \SI{40}{\micro\metre}). The device was operated in continuous wave (CW) at a heat sink temperature of 30 K. In panel (a), we show a free-running fundamental comb spanning around 800 GHz with a single strong RF beatnote at the roundtrip frequency of $f_{\mathrm{rep}}$ = 15.4 GHz. Free-running harmonic comb states\cite{kazakov_self-starting_2017, ForrerAPL2021Harmonic, senica2023frequencymodulated} are observed as well, where the mode spacing is an integer multiple of $f_{\mathrm{rep}}$. A free-running second harmonic comb spanning around 750 GHz is shown in panel (b), with a single RF beatnote at $2f_{\mathrm{rep}}$.

By injecting a strong RF signal into the laser cavity, the emission can be further broadened\cite{senica2022planarized, schneider2021controlling}, as shown in panel (c). However, when the injected RF signal power is too high (typically above 20-30 dBm power at the source, depending on the injection frequency and sample dimensions), this can also produce incoherent (non-comb) states. These can occur due to the limiting chromatic dispersion or due to a voltage swing beyond the threshold/rollover bias point.

\section*{Conclusion}
In conclusion, we have presented a new, high-performance planarized THz quantum cascade laser geometry, where inverse-designed end facet reflectors coupled to surface-emitting antennas result in a seven-fold improvement of the slope efficiency. All the components are optimized for octave-spanning emission spectra, and the measured far-field patterns of broadband devices feature FWHM beam divergences of (17.0\degree x 18.5\degree). The devices operate as broadband frequency combs spanning 800 GHz, with a peak power as high as 13.5 mW. Since the end facet reflectors and antenna are separated by a passive waveguide, we have decoupled the laser mirror reflectivities and the outcoupling structure, allowing for independent control of both aspects.

Moreover, by further reducing the mirror reflectivities and improving the output beam quality, external cavity\cite{hugi2010external, wysocki2005widely} broadband THz QCLs could be fabricated, where the comb repetition rate is continuously tunable by an external mirror/grating. In principle, a near-zero reflectivity end facet can be fabricated by an adiabatic tapered transition between the active and the passive waveguide, facilitating the lasing operation on external cavity modes. In contrast to THz QC-VECSELs \cite{curwen2018terahertz, wu2023rfinjection}, comb operation should be more easily obtained as the planarized ridge waveguide can naturally provide both a broad gain and many longitudinal lasing modes.  

Finally, it is important to emphasize that the presented device is a general layout that can be used with any type of THz QCL active material to engineer the far-field properties and enhance the outcoupling efficiency with a planar and monolithic fabrication technology. In combination with high-power, high-temperature epilayers such as the ones presented in Refs. \cite{bosco_thermoelectrically_2019,khalatpour_high-power_2020}, the peak powers of simple ridge devices could easily go into the Watt-level range while simultaneously being directed into a narrow beam due to the antenna.


 \subsection*{Acknowledgements} The authors gratefully acknowledge funding from the ERC Grant CHIC (grant n. 724344) and in part from the SNF MINT project n. $200021-212735$ and EU project iFLOWS (grant n. 101057844).  
\subsection*{Competing Interests} The authors declare that they have no competing financial interests.
 
\subsection*{Correspondence}  *Correspondence should be addressed to U. Senica (email: usenica@phys.ethz.ch) and G. Scalari (email: scalari@phys.ethz.ch).

\subsection*{Data availability}  All the simulation and experimental data supporting this study are available from the corresponding author upon reasonable request.




\section*{References}\label{References}
\bibliography{GS_bib_PlanarizedWG.bib}

\end{document}